\documentclass[submission, Proceedings]{SciPost}

\hypersetup{
    colorlinks,
    linkcolor={red!50!black},
    citecolor={blue!50!black},
    urlcolor={blue!80!black}
}

\usepackage[bitstream-charter]{mathdesign}
\DeclareSymbolFont{usualmathcal}{OMS}{cmsy}{m}{n}
\DeclareSymbolFontAlphabet{\mathcal}{usualmathcal}

\newcommand{\lsim}{\mathrel{\mathop{\kern 0pt \rlap
  {\raise.2ex\hbox{$<$}}} \lower.9ex\hbox{\kern-.190em $\sim$}}}
\newcommand{\gsim}{\mathrel{\mathop{\kern 0pt \rlap
  {\raise.2ex\hbox{$>$}}}
  \lower.9ex\hbox{\kern-.190em $\sim$}}}

\begin{document}

\begin{center}{\Large \textbf{
Dark Matter:
DAMA/LIBRA and its perspectives\\
}}\end{center}

\begin{center}
R. Bernabei\textsuperscript{1,2},
P. Belli\textsuperscript{1,2$\star$},
F. Cappella\textsuperscript{3,4},
V. Caracciolo\textsuperscript{1,2},
R. Cerulli\textsuperscript{1,2},
C.J. Dai\textsuperscript{5},
A. d'Angelo\textsuperscript{3,4},
A. Incicchitti\textsuperscript{3,4},
A. Leoncini\textsuperscript{1,2},
X.H. Ma\textsuperscript{5},
V. Merlo\textsuperscript{1,2},
F. Montecchia\textsuperscript{2,6},
X.D. Sheng\textsuperscript{5} and
Z.P. Ye\textsuperscript{5,7}
\end{center}

\begin{center}
{\bf 1} Dip. Fisica, Università di Roma ‘‘Tor Vergata’’, 00133 Rome, Italy
\\
{\bf 2} INFN sezione di Roma ‘‘Tor Vergata’’,  00133 Rome, Italy
\\
{\bf 3} Dip. Fisica, Università di Roma ‘‘La Sapienza’’, 00185 Rome, Italy
\\
{\bf 4} INFN sezione di Roma, 00185 Rome, Italy
\\
{\bf 5} Key Laboratory of Particle Astrophysics, Institute of High Energy Physics, \\
Chinese Academy of Sciences, 100049 Beijing, PR China
\\
{\bf 6} Dip. Ingegneria Civile e Ingegneria Informatica, Università di Roma ‘‘Tor Vergata’’, \\
00133 Rome, Italy
\\
{\bf 7} University of Jinggangshan, Ji’an, Jiangxi, PR China
\\

* pierluigi.belli@roma2.infn.it
\end{center}

\begin{center}
\today
\end{center}

\section*{Abstract}
{\bf
The long-standing model-independent annual modulation effect measured by DAMA deep underground at 
Gran Sasso Laboratory with different experimental configurations is summarized and perspectives will be highlighted.
DAMA/LIBRA--phase2 set-up, $\simeq$ 250 kg highly radio-pure NaI(Tl)
confirms the evidence of a signal that meets all the 
requirements of the model independent Dark Matter annual modulation signature at high C.L.; the full exposure is 2.86 ton $\times$ yr over 22 annual cycles.
The experiment is currently collecting data in the DAMA/LIBRA--phase2 empowered configuration with an even lower software energy threshold. 
Other recent claims are shortly commented.}

\section{Introduction}
\label{sec:intro}

The DAMA/LIBRA 
\cite{perflibra,modlibra,modlibra2,modlibra3,review,pmts,mu,norole,daissue,shadow,bot11,mirasim,model,mirsim,uni18,npae18,bled18,npae19,ppnp20,npae21}
experiment,
as the pioneer DAMA/NaI \cite{RNC,ijmd},
has the main aim to investigate the presence of Dark Matter (DM) particles in the galactic halo by exploiting 
the DM annual modulation signature (originally suggested in Ref.~\cite{Freese1,Freese2}). 
In addition, the developed highly radio-pure NaI(Tl) target-detectors 
\cite{perflibra,pmts,daissue,ULBNaI} ensure sensitivity to a wide range 
of DM candidates, interaction types and astrophysical scenarios 
(see e.g. Ref. \cite{ppnp20}, and references therein).
The origin of the DM annual modulation signature and of its peculiar features  is
due to the Earth's revolution around the Sun, 
which is moving in the Galaxy; thus, the Earth should be crossed
by a larger flux of DM particles around $\simeq$ 2 June (when the projection of the Earth orbital velocity on the Sun velocity is maximum)
and by a smaller one around $\simeq$ 2 December (when the two velocities are opposite).
The DM annual modulation signature is very distinctive since the effect
induced by DM particles must simultaneously satisfy
all the following requirements: the rate must contain a component
modulated according to a cosine function (1) with
one year period (2) and a phase that peaks roughly 
$\simeq$ 2 June (3); this modulation must only be found in a
well-defined low energy range, where DM particle induced
events can be present (4); it must apply only to those events
in which just one detector of many actually ``fires'' ({\it single-hit}
 events), since the DM particle multi-interaction probability
is negligible (5); the modulation amplitude in the region
of maximal sensitivity must be $\lsim$ 7\% 
of the constant part of the signal 
for usually adopted halo distributions (6), but it can be larger in case of some
proposed scenarios such as e.g. those reported in Ref.~\cite{ppnp20} 
(even up to $\simeq$ 30\%).
Thus this signature 
has many peculiarities and, in addition, it allows to test a wide range 
of parameters in many possible astrophysical, nuclear and particle physics scenarios.
This DM signature might be mimicked only by systematic effects or side reactions 
able to account for the whole observed modulation amplitude and
to simultaneously satisfy all the requirements given above.

The full description of the DAMA/LIBRA set-up and the adopted procedures 
during the phase1 and phase2 and other related arguments have been discussed in 
details e.g. in Refs. \cite{perflibra,modlibra,modlibra2,modlibra3,review,pmts,uni18,npae18,bled18,npae19,ppnp20,npae21}.

At the end of 2010 all the photomultipliers (PMTs) 
were replaced by a second generation PMTs Hamamatsu R6233MOD, 
with higher quantum efficiency (Q.E.) and with lower background with respect 
to those used in phase1; they were produced  after 
a dedicated R\&D in the company, and tests and selections were described in Refs. \cite{pmts,ULBNaI}. 
The new PMTs have Q.E. in 
the range 33-39\% at 420 nm, wavelength of NaI(Tl) emission, and in the range 36-44\% at peak.
The commissioning of the DAMA/LIBRA--phase2 experiment was successfully performed in 2011, allowing the achievement 
of the software energy threshold at 1 keV, and the improvement of some detector's features 
such as energy resolution and acceptance efficiency near software energy threshold \cite{pmts}. 
The light response of the detectors during phase2 typically ranges 
from 6 to 10 photoelectrons/keV, depending on the detector. 
Energy calibration with X-rays/$\gamma$ sources are regularly carried out in the
same running condition down to few keV (for details see e.g. Ref. \cite{perflibra}; in particular, 
double coincidences due to internal X-rays from $^{40}$K 
(which is at ppt levels in the crystals) provide (when summing the data over long periods)
a calibration point at 3.2 keV close to the software energy threshold. 
The DAQ system records both {\it single-hit} events (where just one of the detectors fires) and 
{\it multiple-hit} events (where more than one detector fires) 
up to the MeV region despite the optimization is performed for the lowest energy.

\section{The DAMA/LIBRA--phase2 results}

The details of the annual cycles of DAMA/LIBRA--phase2 are reported in Ref. \cite{ppnp20,npae21}.
The first annual cycle was dedicated to the commissioning and to the optimizations towards the 
achievement of the 1 keV software energy threshold \cite{pmts}. Thus, the considered annual cycles of DAMA/LIBRA--phase2 released so far
are eight 
(exposure of 1.53 ton$\times$yr); when considering also the former DAMA/NaI and DAMA/LIBRA--phase1, the exposure is 2.86 ton$\times$yr.
The duty cycle of the DAMA/LIBRA--phase2 experiment is high, ranging between 76\% and 86\%.
The routine calibrations and, in particular, the data collection for the acceptance windows 
efficiency mainly affect it. 

\begin{figure}[!ht]
\begin{center}
\vspace{-0.4cm}
\includegraphics[width=\textwidth] {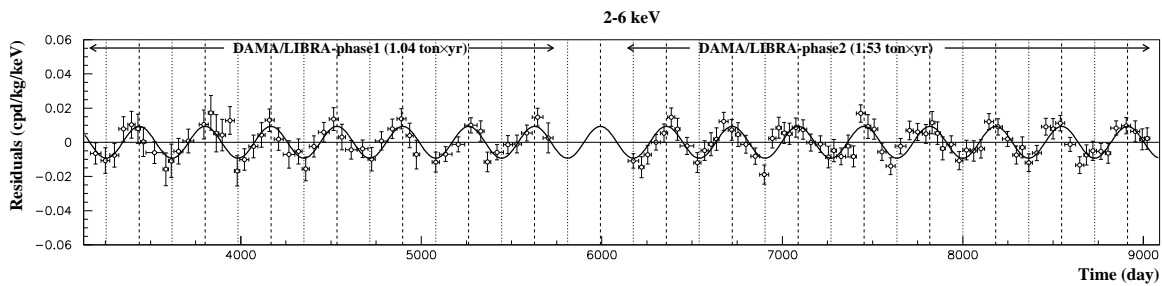}
\end{center}
\vspace{-0.6cm}
\caption{Experimental residual rate of the {\it single-hit} scintillation events
measured by DAMA/LIBRA--phase1 and DAMA/LIBRA--phase2 
in the (2--6) keV energy intervals
as a function of the time. 
The superimposed curve is the cosinusoidal functional forms $A \cos \omega(t-t_0)$
with a period $T = \frac{2\pi}{\omega} =  1$ yr, a phase $t_0 = 152.5$ day (June 2$^{nd}$) and
modulation amplitude, $A$, equal to the central value obtained by best fit. This figure is being reused from \cite{npae21}.}
\label{fg:res2}
\end{figure}

Residual rates versus time for 1 keV energy threshold are reported in Ref. \cite{npae21}.
The former DAMA/LIBRA--phase1 and the new DAMA/LIBRA--phase2 residual rates of the {\it single-hit} scintillation events are
reported in Fig.~\ref{fg:res2}. The energy interval is from 2 keV, the software energy threshold of DAMA/LIBRA--phase1,
up to 6 keV. The data of Fig.~\ref{fg:res2} and those of DAMA/NaI
have been fitted with the function: $A \cos \omega(t-t_0)$, considering a
period $T = \frac{2\pi}{\omega} =  1$ yr and a phase $t_0 = 152.5$ day (June 2$^{nd}$) as 
expected by the DM annual modulation signature.
The obtained $\chi^2/d.o.f.$ is 130/155 and the modulation amplitude $A=(0.00996 \pm 0.00074)$ cpd/kg/keV is obtained.
When the period and the phase are kept free in the fitting procedure,
the achieved C.L. for the full exposure (2.86 ton$\times$yr) is  $13.7 \sigma$;
the modulation amplitude of the {\it single-hit} scintillation events 
is: $(0.01014 \pm 0.00074)$ cpd/kg/keV,
the measured phase is $(142.4 \pm 4.2)$ days 
and the measured period is $(0.99834 \pm 0.00067)$ yr, 
all these values are well in agreement with those expected for DM particles. 

Absence of any significant 
background modulation in the energy spectrum has also been verified 
in the present data taking for energy regions not of interest for DM \cite{modlibra,modlibra2,modlibra3,review,daissue,uni18,npae18,bled18,ppnp20,npae21}.
It is worth noting that the obtained results account of whatever kind of background and, in addition, 
no background process able to mimic the DM annual modulation signature (that is able to simultaneously satisfy 
all the peculiarities of the signature and to account for the measured modulation amplitude) is available 
(see also discussions e.g. in Ref.~\cite{perflibra,modlibra,modlibra2,modlibra3,review,mu,norole,uni18,npae18,bled18,ppnp20,npae21}).

A further relevant investigation on DAMA/LIBRA--phase2 data has been performed by 
applying the same hardware and software 
procedures, used to acquire and to analyze the {\it single-hit} residual rate, to the 
{\it multiple-hit} one. 
Since the 
probability that a DM particle interacts in more than one detector 
is negligible, a DM signal can be present just in the {\it single-hit} residual rate.
Thus, the comparison of the results of the {\it single-hit} events with those of the  {\it 
multiple-hit} ones corresponds to compare the cases of DM particles beam-on 
and beam-off.
This procedure also allows an additional test of the background behaviour in the same energy interval 
where the positive effect is observed. 
While a clear modulation, satisfying all the peculiarities of the DM
annual modulation signature, is present in the {\it single-hit} events,
the fitted modulation amplitude for the {\it multiple-hit}
residual rate is well compatible with zero \cite{npae21}.
Since the same identical hardware and the same identical software procedures have been used to 
analyze the two classes of events, the obtained result offers an additional strong support for the 
presence of a DM particle component in the galactic halo.

The {\it single-hit} residuals have also been investigated by a Fourier analysis \cite{review}. A clear peak corresponding to a period of 1 year is evident in the low energy intervals; the same analysis 
in the (6–14) keV energy region shows only aliasing peaks instead. Neither other structure at different frequencies has been observed.

\begin{figure}[!h]
\begin{center}
\vspace{-0.4cm}
\includegraphics[width=0.7\textwidth] {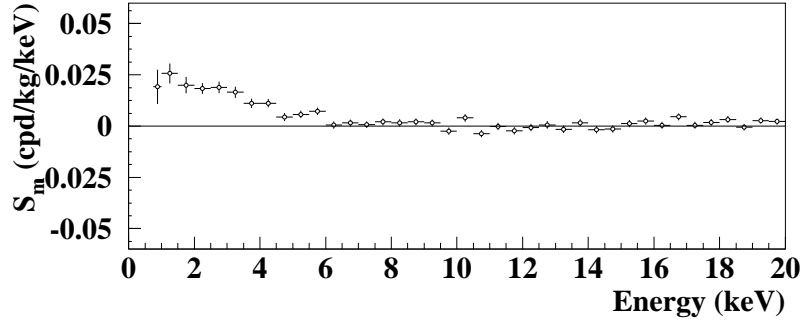}
\end{center}
\vspace{-0.6cm}
\caption{Modulation amplitudes, $S_{m}$, as function of the energy in keV(ee) for the whole data sets: DAMA/NaI, DAMA/LIBRA--phase1 and DAMA/LIBRA--phase2 
(total exposure 2.86 ton$\times$yr) above 2 keV; below 2 keV only the DAMA/LIBRA-phase2 exposure (1.53 ton $\times$ yr) is available and used. 
A clear modulation is present in the lowest energy region,
while $S_{m}$ values compatible with zero are present just above. This figure is being reused from \cite{npae21}.
}
\label{fg:sme}
\end{figure}

The annual modulation present at low energy can also 
be pointed out by depicting the energy dependence of
the modulation amplitude, $S_{m}(E)$, obtained
by maximum likelihood method considering fixed period and phase: $T=$1 yr and $t_0=$ 152.5 day.
The modulation amplitudes for the whole data sets: DAMA/NaI, DAMA/LIBRA--phase1 and DAMA/LIBRA--phase2 
(total exposure 2.86 ton$\times$yr) are plotted in Fig.~\ref{fg:sme};  
the data below 2 keV refer only to the 
DAMA/LIBRA-phase2 exposure (1.53 ton$\times$yr).
It can be inferred that positive signal is present in the (1--6) keV energy interval (a new data point below 1 keV has been added, see later), while $S_{m}$
values compatible with zero are present just above. 
All this confirms the previous analyses. 
The test of the hypothesis that the $S_{m}$ values
in the (6--14) keV energy interval have random fluctuations around zero 
yields $\chi^2/d.o.f.$ equal to 20.3/16 (P-value = 21\%).

It has been verified that the observed annual modulation effect is well distributed in all the 25 detectors. 
In particular, the modulation amplitudes $S_{m}$ integrated in the range (2--6) keV for each of the 25 detectors
for the DAMA/LIBRA--phase1 and DAMA/LIBRA--phase2 periods have random fluctuations around the weighted averaged value 
confirmed by the $\chi^2$  analysis.
Thus, the hypothesis that the signal is well distributed over all the 25 detectors is accepted.

Among further additional tests, the analysis 
of the modulation amplitudes separately for
the nine inner detectors and the external ones has been carried out for DAMA/LIBRA--phase1 and DAMA/LIBRA--phase2,
as already done for the other data sets \cite{modlibra,modlibra2,modlibra3,review,uni18,npae18,bled18,ppnp20,npae21}. 
The obtained values are fully in agreement; in fact,
the hypothesis that the two sets of modulation amplitudes belong to same distribution has been verified by $\chi^2$ test, obtaining e.g.:
$\chi^2/d.o.f.$ = 1.9/6 and 
36.1/38 for the energy intervals (1--4) and (1--20) keV, 
respectively ($\Delta E$ = 0.5 keV). This shows that the
effect is also well shared between inner and outer detectors. 

To test the hypothesis that the modulation amplitudes calculated for each DAMA/LIBRA--phase1 and DAMA/LIBRA--phase2 
annual cycle are compatible and normally fluctuating around their mean values, the $\chi^2$ test and the {\it run test} have been used. 
This analysis confirms that the data collected in all the annual cycles with DAMA/LIBRA--phase1 and phase2 are statistically compatible
and can be considered together \cite{npae21}.

Let us, finally, release the assumption of the phase $t_0=152.5$ day in the procedure to 
evaluate the modulation amplitudes. In this case the signal can be alternatively written as:
\begin{eqnarray}
\label{eqn1} 
S_{i}(E) & = & S_{0}(E) + S_{m}(E) \cos\omega(t_i-t_0) + Z_{m}(E) \sin\omega(t_i-t_0) \\
        & = & S_{0}(E) + Y_{m}(E) \cos\omega(t_i-t^*).   \nonumber
\end{eqnarray}

\noindent For signals induced by DM particles one should expect: 
i) $Z_{m} \sim 0$ (because of the orthogonality between the cosine and the sine functions); 
ii) $S_{m} \simeq Y_{m}$; iii) $t^* \simeq t_0=152.5$ day. 
These conditions hold for most of the dark halo models; however,
slight differences can be expected in case of possible contributions
from non-thermalized DM components (see e.g. Ref. \cite{ppnp20} and references therein).

\begin{figure}[!ht]
\begin{center}
\includegraphics[width=0.35\textwidth] {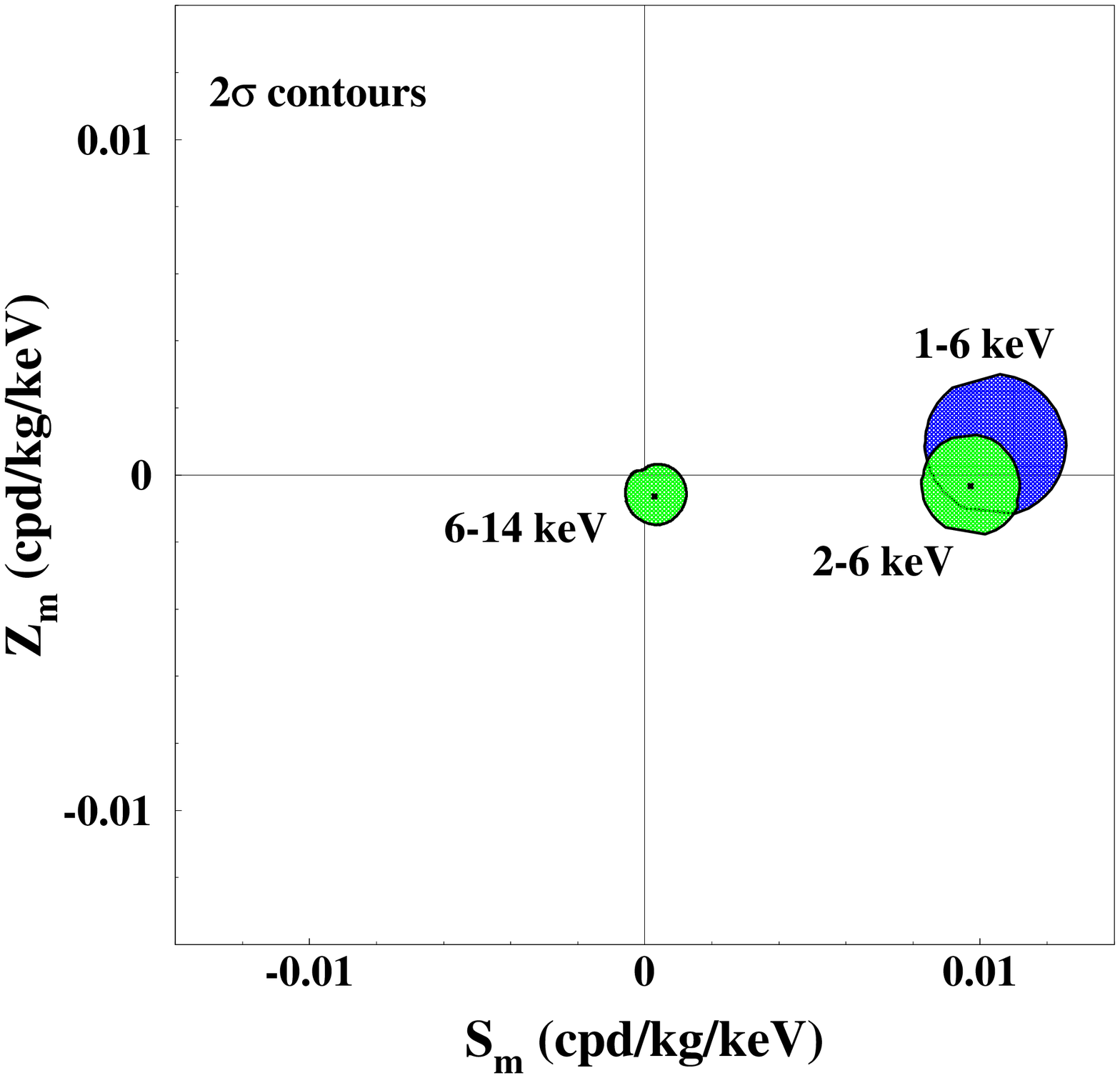}
\includegraphics[width=0.35\textwidth] {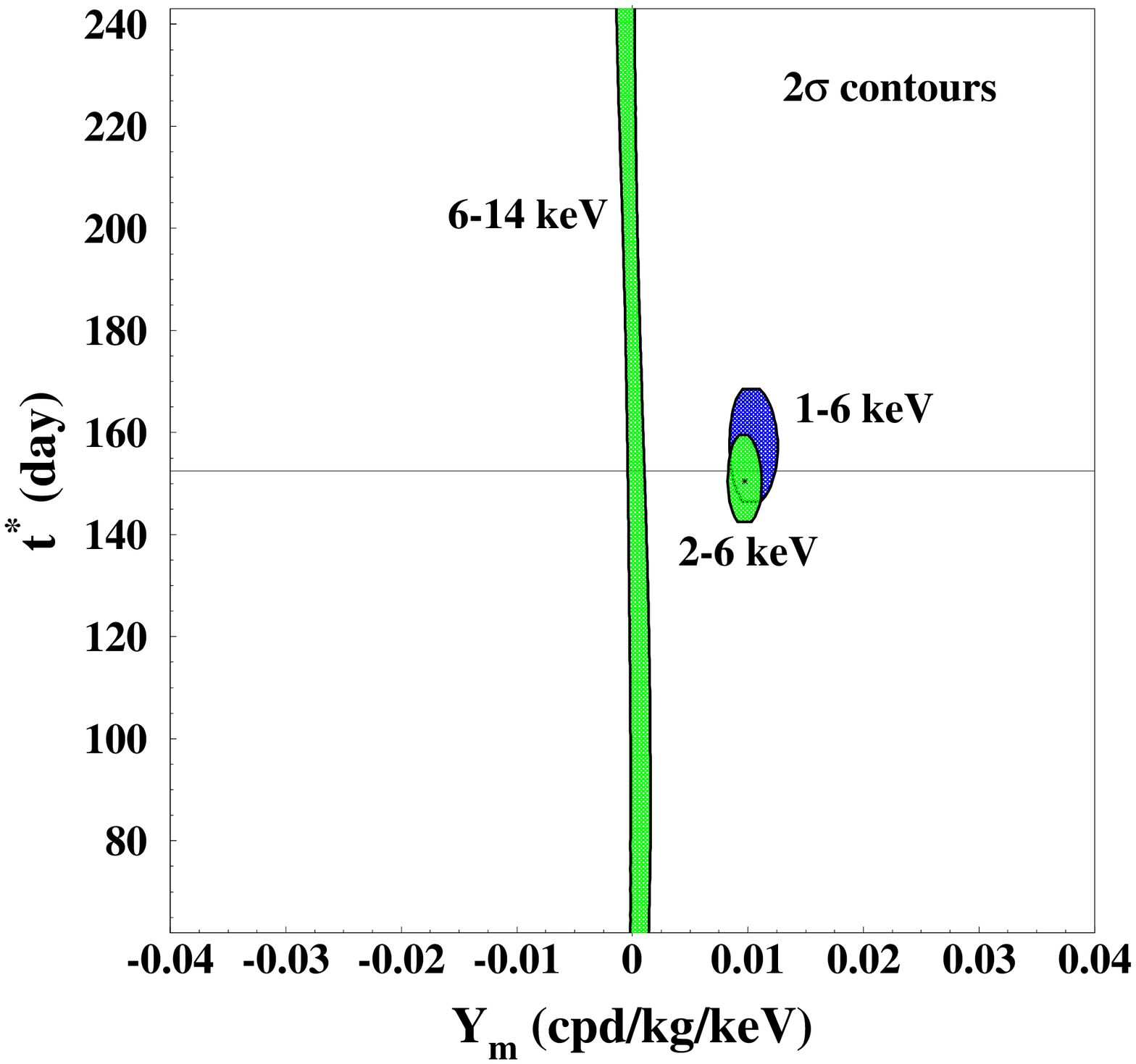}
\end{center}
\vspace{-0.8cm}
\caption{
$2\sigma$ contours in the plane $(S_m , Z_m)$ ({\it left})
and in the plane $(Y_m , t^*)$ ({\it right})
for:
i) DAMA/NaI, DAMA/LIBRA--phase1 and DAMA/LIBRA--phase2 in the (2--6) keV and (6--14) keV energy intervals (light areas, green on-line);
ii) only DAMA/LIBRA--phase2 in the (1--6) keV energy interval (dark areas, blue on-line).
The contours have been  
obtained by the maximum likelihood method.
A modulation amplitude is present in the lower energy intervals 
and the phase agrees with that expected for DM induced signals. These figures are being reused from \cite{npae21}.
}
\label{fg:bid}
\end{figure}

Considering cumulatively the data of DAMA/NaI, DAMA/LIBRA--phase1 and DAMA/LIBRA--phase2
the obtained $2\sigma$ contours in the plane $(S_m , Z_m)$ 
for the (2--6) keV and (6--14) keV energy intervals 
are shown in Fig.~\ref{fg:bid}{\it --left} while in 
Fig.~\ref{fg:bid}{\it --right} the obtained $2\sigma$ contours in the plane $(Y_m , t^*)$
are depicted.
Moreover, Fig.~\ref{fg:bid} also shows  only for DAMA/LIBRA--phase2 
the $2\sigma$ contours in the (1--6) keV energy interval.
The best fit values are reported in Ref. \cite{npae21}.

Setting $S_{m}=0$ in eq.~(\ref{eqn1}), 
the $Z_{m}$ values have also been determined
by using the same procedure for
DAMA/NaI, DAMA/LIBRA--phase1 and phase2 data sets; they
are expected to be zero. 
The $\chi^2$ test supports the hypothesis that the $Z_{m}$ values are simply 
fluctuating around zero; in fact, 
in the (1--20) keV energy region the $\chi^2/d.o.f.$
is equal to 40.6/38 corresponding to a P-value = 36\%.

No systematic or side processes able to mimic the signature, i.e. able to 
simultaneously satisfy all the many peculiarities 
of the signature and to account for the whole measured modulation amplitude, has been 
found or suggested by anyone throughout some decades thus far
(for details see e.g. Ref. \cite{perflibra,modlibra,modlibra2,modlibra3,review,mu,norole,RNC,ijmd,uni18,npae18,npae19,bled18,ppnp20,npae21}).

In particular, arguments related to any possible role of some natural 
periodical phenomena have been discussed and quantitatively demonstrated 
to be unable to mimic the signature (see references; e.g. Refs. \cite{mu,norole}).
Thus, on the basis of the exploited signature, the model independent DAMA results give evidence 
at 13.7$\sigma$ C.L. (over 22 independent annual cycles and in various experimental configurations) for 
the presence of DM particles in the galactic halo.

The DAMA model independent evidence is compatible with a wide 
set of astrophysical, nuclear and particle physics scenarios 
for high and low mass candidates inducing nuclear recoil and/or electromagnetic radiation,
as also shown in various literature.
Moreover, both the negative results and all the possible positive hints, achieved so-far
in the field, can be compatible with the DAMA model independent DM annual modulation results in many scenarios considering also the 
existing experimental and theoretical uncertainties; the same holds for indirect approaches. 
For a discussion see e.g. Ref.~\cite{review,ppnp20} and references therein.

\section{Few arguments about the analysis procedure}

As reported several times along the years \cite{modlibra,modlibra2,modlibra3,review,uni18,npae18,bled18,ppnp20,npae21},
the data taking of each annual cycle in DAMA/LIBRA starts before the expected minimum of the DM signal (about 2 December) 
and ends after its expected maximum (about 2 June). 
Thus, adopting in the data analysis a constant background evaluated within each annual cycle, 
any possible decay of long--term--living isotopes cannot mimic a DM positive signal with all its peculiarities.
On the contrary, it may only lead to underestimate the DM annual modulation amplitude, depending on the radio-purity of the set-up.

Despite this obvious fact, Refs. \cite{panci20,cos22} claim that the DAMA annual modulation result 
might be mimicked by the adopted analysis procedure. Detailed analyses on this argument have already been reported in Ref. \cite{ppnp20},
confuting these claims quantitatively, even considering the case of a rate at low energy in DAMA/LIBRA with odd behavior, increasing with time.

More recently, Ref. \cite{cos22} claims that an annual modulation 
in the COSINE--100 data can appear if they use an analysis method somehow similar to DAMA/LIBRA.
However, as expected from the rate of COSINE--100 very--decreasing with time and from what mentioned above, the authors obtain
a modulation with reverse phase \cite{cos22}; this corresponds, when fixing the phase to $t_0=152.5$ day, to
{\it NEGATIVE} modulation amplitudes, as expected by the elementary considerations reported before. This artificial effect has no way to mimic the observed DM signature with its peculiarity.

Thus, while the appearance of modulation with  {\it NEGATIVE} amplitudes is due to the peculiar behavior of the COSINE--100 rate very--decreasing with time, 
this is not the case of DAMA/LIBRA. In particular, the DAMA/LIBRA NaI(Tl) detectors are not the ‘‘same’’ as those of COSINE--100, since e.g. they were grown starting 
from different powders, using different purification, growing procedures and protocols; they have been stored underground since decades, 
they have different quenching factors for alpha's and nuclear recoils, etc. Thus, they have well different residual contaminations and features\footnote{
The DAMA/LIBRA set-up had some upgrades -- one of them is that from phase1 to phase2 to lower the software energy threshold -- also acting to improve the signal/background ratio.} 
as well as different electronics and all other details of the experimental set-up.

Moreover, the stability with time of the running parameters of each DAMA/LIBRA annual cycle is reported e.g. in Refs. \cite{modlibra,modlibra2,modlibra3,review,uni18,npae18,bled18,ppnp20,npae21}.
As regards the odd idea that the low-energy rate in DAMA/LIBRA might increase with time due to spill out of noise \cite{cos22} , we just recall two facts  
that rule out this possibility: 1) the stability with time of noise, reported in several papers \cite{modlibra,modlibra2,modlibra3,review,uni18,npae18,bled18,ppnp20,npae21}; 
2) the estimate of the remaining noise tail after the noise rejection procedure $\ll 1\%$ \cite{pmts}. 

Finally,  the arguments of Ref. \cite{ppnp20} already showed that any possible effect in DAMA/LIBRA due to either long--term time--varying background 
or odd behavior of the rate, increasing with time, is negligible. Here we just recall:

\begin{itemize}

\item The (2--6) keV {\it single-hit} residual rates have been recalculated considering a possible time--varying background. They provide 
modulation amplitude, fitted period and phase well compatible with those obtained in the  {\it original}  analysis, showing that the
effect of long--term time--varying background -- if any -- is marginal \cite{ppnp20}.

\item Any possible long--term time--varying background would also induce a fake modulation amplitudes ($\Sigma$) on the
tail of the $\mathcal{S}_m$ distribution above the energy region where the signal has been observed. The analysis in Ref. \cite{ppnp20}
shows that $\mid \Sigma \mid < 1.5 \times 10^{-3}$ cpd/kg/keV. Thus, taking into account that the observed {\it single-hit} annual modulation amplitude at low energy is order of $10^{-2}$
cpd/kg/keV,  any possible effect of long--term time--varying background -- if any -- is marginal \cite{ppnp20}.

\item The maximum likelihood analysis has been repeated including
a linear term decreasing with time. The obtained $\mathcal{S}_m$ averaged over the low energy interval are compared with those obtained  in the  {\it original}  analysis,
showing that their differences are well below the statistical errors \cite{ppnp20}.

\item The behaviour of the {\it multiple-hit} events, where no modulation has been found \cite{ppnp20,npae21} in the same energy region where the 
annual modulation is present in the {\it single-hit} events, strongly disfavours the hypothesis that the counting rate has significant long--term time--varying contributions. 

\end{itemize}

Summarizing, the arguments of Ref. \cite{ppnp20} already showed that any possible effect in DAMA/ LIBRA due either to long--term time--varying background 
or to any odd behavior of the rate, increasing with time, is negligible and the {\it original} analyses, that assume a constant background within each annual cycle, can be safely adopted.
Similar conclusions were also reported in Ref. \cite{mess20}.

\section{Perspectives, comparisons and conclusions}

To further increase the experimental sensitivity of DAMA/LIBRA and to disentangle some of the many possible astrophysical, nuclear and particle physics scenarios in the investigation on the DM candidate particle(s), 
an increase of the exposure in the lowest energy bin and a further decreasing of the software energy threshold are needed. This is pursued by running DAMA/LIBRA--phase2 and 
upgrading the experimental set-up to lower the software energy threshold below 1 keV with high acceptance efficiency.

Firstly, particular efforts for lowering the software energy threshold have been done in the already-acquired data of DAMA/LIBRA--phase2 by using the same technique as before with dedicated studies on the 
efficiencies. Consequently, a new data point has been added in the modulation amplitude as a function of energy down to 0.75 keV, see Fig.~\ref{fg:sme}.
A modulation is also present below 1 keV. This preliminary result confirms the necessity to lower the software energy threshold by a hardware upgrade 
and an improved statistics in the first energy bin.

A dedicated hardware upgrade of DAMA/LIBRA–phase2 was done. All the PMTs were equipped with miniaturized low background new concept preamplifiers and miniaturized HV dividers mounted 
on the same socket. The electronic chain was improved mainly by using higher vertical resolution 14--bit digitizers. This upgrade aims to improve the experimental sensitivity through 
a lower software energy threshold and a large acceptance efficiency.
The experiment is currently running in this new configuration, DAMA/LIBRA--phase2 empowered, and new results are foreseen in the near future.

\end{document}